\documentclass[12pt,%
 reprint,
 amsmath,amssymb,
 aps,
]{aastex63}

\usepackage{graphicx}
\usepackage{dcolumn}
\usepackage{bm}
\usepackage{multirow}
\newcommand\pp{\bf}
\newcommand{\add}{}
\usepackage{soul}

\begin{document}


\title{Role of Parallel Solenoidal Electric Field on Energy Conversion in 2.5D Decaying Turbulence with a Guide Magnetic Field}

\author{Peera Pongkitiwanichakul}
 \email{peera.po@ku.th}
\affiliation{Department of Physics, Faculty of Science, Kasetsart University, Bangkok 10900, Thailand}

\author{David Ruffolo}
\affiliation{%
Department of Physics, Faculty of Science, Mahidol University, Bangkok 10400, Thailand
}%

\author{Fan Guo}%
\affiliation{%
Los Alamos National Laboratory, Los Alamos, NM 87545, USA
}
\affiliation{%
New Mexico Consortium, Los Alamos, NM 87544, USA
}%

\author{Senbei Du}
\affiliation{%
Los Alamos National Laboratory, Los Alamos, NM 87545, USA
}%

\author{Piyawat Suetrong}
\affiliation{%
Department of Physics, Faculty of Science, Kasetsart University, Bangkok 10900, Thailand
}%

\author{Chutima Yannawa}
\affiliation{%
Department of Physics, Faculty of Science, Kasetsart University, Bangkok 10900, Thailand
}%

\author{Kirit Makwana}
\affiliation{Indian Institute of Technology Hyderabad
Kandi-502285, Sangareddy
Telangana, India 
}%

\author{Kittipat Malakit}
\affiliation{%
Department of Physics, Faculty of Science and Technology, Thammasat University, Pathum Thani 12120, Thailand
}%

\date{\today}

\begin{abstract}
We perform 2.5D particle-in-cell simulations of decaying turbulence in the presence of a guide (out-of-plane) background magnetic field.
The fluctuating magnetic field initially consists of Fourier modes at low wavenumbers (long wavelengths). 
With time, the electromagnetic energy is converted to plasma kinetic energy (bulk flow+thermal energy) at the rate per unit volume of ${\pp J}\cdot{\pp E}$ for current density ${\pp J}$ and electric field ${\pp E}$. Such decaying turbulence is well known to evolve toward a state with strongly intermittent plasma current. Here we decompose the electric field into components that are irrotational, ${\pp E}_{\rm ir}$, and solenoidal (divergence-free), ${\pp E}_{\rm so}$. ${\pp E}_{\rm ir}$ is associated with charge separation, and ${\pp J}\cdot{\pp E}_{\rm ir}$ is a rate of energy transfer between ions and electrons with little net change in plasma kinetic energy. Therefore, the net rate of conversion of electromagnetic energy to plasma kinetic energy is strongly dominated by ${\pp J}\cdot{\pp E}_{\rm so}$, and for a strong guide magnetic field, this mainly involves the component ${\pp E}_{\rm so,\parallel}$ parallel to the total magnetic field ${\pp B}$. 
We examine various indicators of the spatial distribution of
the energy transfer rate {\bf J$_\parallel\cdot$E$_{so,\parallel}$}, which relates to magnetic reconnection, the best of which are  1) the ratio of the out-of-plane electric field to the in-plane magnetic field, 2) the out-of-plane component of the non-ideal electric field, and 3) the magnitude of the estimate of current helicity.

\end{abstract}


\section{\label{intro} Introduction}

The conversion of magnetic field energy into internal energy of plasmas in weakly collisional or collisionless plasmas is not fully understood and has important implications for space plasma and astrophysical plasma phenomena~\citep{parashar2015}. 
A magnetic field in a non-potential form can serve as a source of energy for plasma heating and/or particle acceleration in numerous space, solar and astrophysical processes such as coronal heating~\citep{ballegooijen86,parker88,peera2015}, solar flares~\citep{li2017,Chen2020,Fu2020}, solar wind acceleration~\citep{pontieu07}, and flares in relativistic jets \citep{Zhang2015,Guo2015,Comisso2018,Zhang2018,Zhang2020}. These events involve multiple phenomena operating on a wide range of scales. At large scales, the plasma behaves like a fluid and nonlinear interactions such as turbulence can cause an energy cascade toward small scales where kinetic physics becomes important~\citep{leamon98,gary2004,sahraoui2009,chandran10,wan2012,perri2012,tenBarge2013}. While the overall energy conversion rate can be controlled by turbulence at large scales~\citep{wu13,peera2015}, unclear mechanisms at kinetic scales determine the proportion of energy transferred to ions or electrons. Ions gain much more energy than electrons in  high-energy turbulent astrophysical systems~\citep{zhdankin2019}.  
Changing the level of fluctuation at kinetic scales causes a change in dominant heating mode. For a higher amplitude of fluctuation, ions gain energy more effectively, while a lower amplitude supports more electron heating~\citep{gary16,hughes17,shay2018}.

A number of recent works searched for the responsible mechanisms and the locations for the energy conversion by using the statistics of properties of the energy transfer rate per unit volume from electromagnetic energy to plasma energy, ${\pp J}\cdot{\pp E}$, where ${\pp J}$ is a current density and ${\pp E}$ is an electric field.
Some previous studies explored how different components of ${\pp J}\cdot{\pp E}$ contribute to the total conversion rate. 
For example, one can consider the perpendicular component of ${\pp J}$ as the sum of multiple currents due to the anisotropic and non-uniform pressure tensor, such as the grad-B current, the curvature current, and the polarization current~\citep{Dahlin2014,li2017,li2018,Li2019}. In the case of large-scale magnetic reconnection, ${\pp J}\cdot{\pp E}$ from the curvature current dominates. In the case of decaying turbulence with no dominant magnetic reconnection site, the currents are also intermittent and form sheet-like structures~\citep{camporeale2018}.
For this case, the main contribution to  ${\pp J}\cdot{\pp E}$  comes from the interactions between particles and the parallel component of the electric field~\citep{wan2012,makwana2017}. Another line of work has considered the conversion into internal energy due to the interaction between the pressure tensor and gradient of the components of bulk flows \citep{Yang2017,Yang2017PRE,Du2018,Du2020}.

\begin{figure}[ht]
  \begin{center}
    \includegraphics[width=0.7\textwidth]{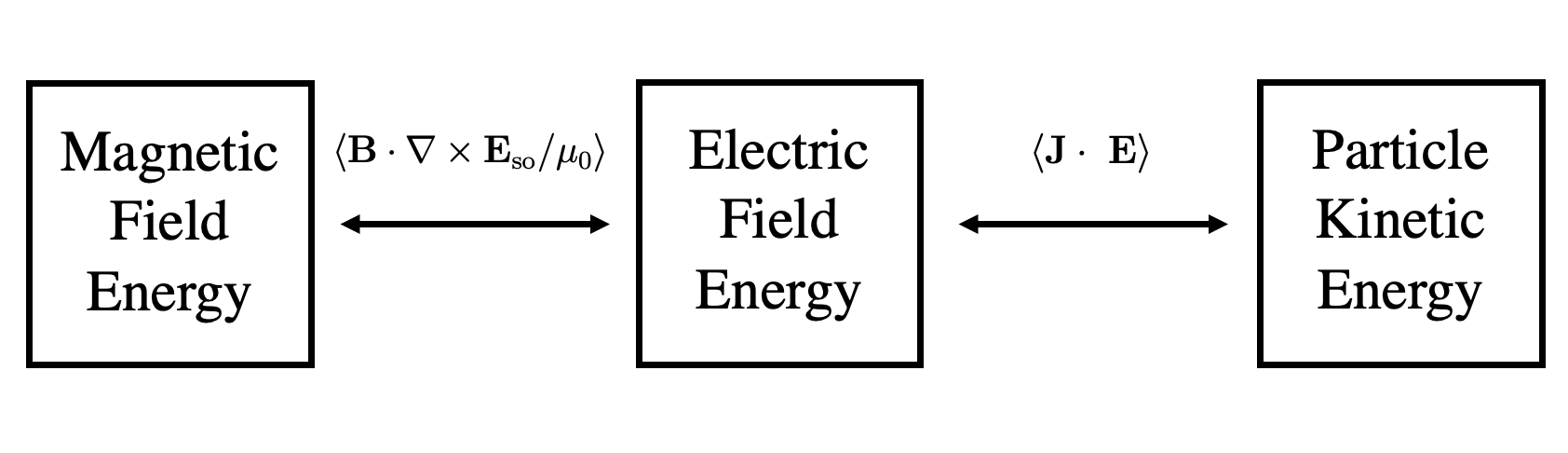}
    \caption{\small Schematic of the net energy transfer between the magnetic field, the electric field, and particles via $\langle {\pp B}\cdot\nabla\times{\pp E}_{\rm so}/\mu_0\rangle$ and $\langle {\pp J}\cdot{\pp E}\rangle$. According to our simulation results, for decaying 2.5D turbulence the two net energy transfer rates are nearly equal, and the net transfer $\langle{\bf J\cdot E}\rangle$ is dominated by the solenoidal component $\langle{\bf J\cdot E_{so}}\rangle$. }
\label{fig:conversion}
  \end{center}
\end{figure}

Examination of the spatial distribution, species contributions, and directional components of ${\pp J}\cdot{\pp E}$ is useful for understanding how a magnetic field loses its energy to particles.
Furthermore, we point out the utility of decomposing ${\pp E}$ into irrotational (${\pp E}_{\rm ir}$) and solenoidal (divergence-free, ${\pp E}_{\rm so}$) components. 
The irrotational component arises from the charge separation between ions and electrons. 
Consider the rates of change of magnetic field and electric field energy according to Maxwell's equations:
\begin{eqnarray}
\frac{\partial}{\partial t}\left(\frac{B^2}{2\mu_0}\right) &=& -\frac{1}{\mu_0}{\bf B}\cdot\nabla\times{\bf E} \label{eq:B2} \\ 
\frac{\partial}{\partial t}\left(\frac{\epsilon_0E^2}{2}\right) &=& \frac{1}{\mu_0}{\bf E}\cdot\nabla\times{\bf B}-{\bf J}\cdot{\bf E}. \label{eq:E2}
\end{eqnarray}
The first term on the right hand side of Equation~\ref{eq:E2} has the same spatial average as the negative of the right hand side of Equation~\ref{eq:B2}, which represents the rate of conversion of magnetic energy to electric energy.
From Equation~\ref{eq:B2}, the irrotational component does not contribute to the time evolution of the magnetic field. In that sense, it cannot exchange energy with the magnetic field.
Only the solenoidal component ${\pp E}_{\rm so}$ can directly drain energy from the magnetic field. 
Figure~\ref{fig:conversion} shows an illustration of how overall energy is converted between the magnetic field, electric field and particle kinetic energy (of both bulk flow and thermal energy).
The energy density conversion rate between the magnetic field and the electric field is ${\pp B}\cdot\nabla\times{\pp E}_{\rm so}/\mu_0$, while ${\pp J}\cdot{\pp E}$ is the energy density conversion rate between the electric field and particles. The characteristics of ${\pp E}_{\rm so}$ control the energy conversion between the magnetic field and the electric field. As will be discussed shortly, our simulation results show that the energy density in the electric field remains low with a very slow rate of change, compared to the magnetic energy or kinetic energy.

Furthermore, we find that the spatial average $\langle{\bf J\cdot E_{ir}}\rangle$ is very low.  Therefore, the net energy flow is from the magnetic field through the solenoidal electric field energy and then, at about the same rate, through to the particle kinetic energy via ${\bf J\cdot E_{so}}$. In other words, ${\bf E_{so}}$ controls the flow of magnetic energy to particle kinetic energy, with a spatial distribution marked by ${\bf J\cdot E_{so}}$.
The analysis of ${\pp J}\cdot{\pp E}$ from previous studies generally included a major contribution from ${\pp E}_{\rm ir}$, which may obscure the net energy-conversion mechanism and spatial distribution.

In this work, we investigate how the magnetic energy is converted into the kinetic energy of particles. We run particle-in-cell simulations for cases of 2.5-dimensional (2.5D) decaying turbulence with an initially uniform guide magnetic field.
The purpose of studying decaying turbulence is to study the evolution of turbulence without the strong influence of the turbulent energy input (forcing) that would have to be added {\it ad hoc} to maintain a steady state.  Simulations in 2.5D are useful for effectively utilizing computational resources, particularly for the case of a strong guide field, where the turbulent dynamics involve stronger spatial variations along the two directions perpendicular to that field \citep{montgomery82}, i.e., they have a quasi-2D nature.
We analyze the behavior of ${\pp J}\cdot{\pp E}$ from ions and electrons separately.  We \add{use Helmholtz
decomposition~\citep{kida1992,yang2021} to} separate the electric field into the irrotational component and the solenoidal component.
We focus on how the solenoidal component drains energy out of the magnetic field and find which types of particles gain energy from this component.
We consider multiple magnitudes of the guide field and observe how the energy conversion is modified. 
We find that the simulations have two phases. In the early phase, magnetic islands are formed. In the later phase, these magnetic islands merge. As plasma dynamics in the later phase is less sensitive to the initial conditions and more closely related to astrophysical applications, we focus on the energy transfer in this phase. In the later phase, the interaction between particles and the parallel component of the solenoidal electric field is the main mechanism to drain energy out of the magnetic field in the case of a strong guide field. The 
interaction is strongly localized, and we find that three good indicators of the spatial locations are 1) the ratio of the out-of-plane electric field to the in-plane magnetic field, 2) the out-of-plane component of the non-ideal electric field, and 3) the magnitude of the estimate of current helicity.

\section{\label{model} Simulation Setup}

Simulations presented here were performed using VPIC~\citep{bowers2008}, which is an explicit particle-in-cell (PIC) code that has been used to study magnetic reconnection \citep{daughton2011,Guo2020a,Guo2020b}, turbulence \citep{wan2015}, and particle energization \citep{Guo2014,Guo2015,Guo2019,Li2019b}. 
The simulations are 2.5D with $y$  as the ignorable coordinate. The initial magnetic field is the sum of a uniform guide magnetic field and an additional in-plane magnetic field in the form of several Fourier modes with a certain range of wavenumbers:

\begin{equation}
    {\pp B}(x, z) = B_0\hat{y}+ {\pp \delta b} (x, z),
\end{equation}
where
\begin{equation}
    {\pp \delta b}(x,z)= \sum_{n=1}^{8}\sum_{m=1}^{8} {\pp b}(n,m)e^{i(2\pi n x/L+2\pi m z/L+\phi_{\rm mn})},
\end{equation}
$B_0\hat{y}$ is the uniform guide field and $L$ is the dimension of the cubical box. The wavenumber components along the $x$ and $z$ directions vary from $2\pi/L$ to $16\pi/L$. The phases $\phi_{\rm mn}$ are randomized.
The Fourier coefficient ${\pp b}(n,m)$ is initially in the $x$-$z$ plane and is perpendicular to the corresponding wavenumber to maintain $\nabla\cdot {\pp B}=0$. The magnitude of ${\pp b}(n,m)$ is uniform and is set to give $\langle {\pp \delta b}^2\rangle=b_0^2$, where $b_0$ is the unit of the magnetic field in the simulations.
This form of the magnetic field is associated with plasma current along the $+y$ and $-y$ directions and contains a total energy available for energy conversion of equal to $(\langle B^2 \rangle-B^2_0)L^3/(2\mu_0)=b^2_0 L^3/(2\mu_0)$. There is no initial electric field and the plasma pressure is uniform. Therefore at its early state, the forces in the plasma are extremely unbalanced.

The plasma consists of ion-electron pairs with the ion-to-electron mass ratio $m_i/m_e$ of $25$ or $100$. The initial particle distributions are Maxwellian with a uniform density $n_0$ and a uniform temperature for ions and electrons.
The initial electron thermal speed is $0.2c$.  
The initial bulk flow velocities of ions and electrons give a net-zero mass flow but provide the current density ${\pp J}=\nabla\times{\pp B}/\mu_0$.  The unit of time is the nominal inverse ion cyclotron frequency for the in-plane magnetic field, $\Omega^{-1}_{\rm i}=m_{\rm i}/(q_{\rm i} b_0)$ and the unit of length is the initial ion inertial length $d_i = c/\omega_{\rm pi}$, where $\omega_{\rm pi}$ is the initial ion plasma frequency. The initial density $n_0$
is set so that the initial electron plasma frequency
is $\omega_{\rm pe}=\Omega_{\rm e}=q_{\rm e}b_0/m_{\rm e}$.

For $m_i/m_e=25$, we ran simulations with $B_0/b_0$ =$0.1$, $0.5$, $1$ and $2$. For $m_i/m_e=100$, the simulation is  with $B_0/b_0=2$. Therefore, the initial values of plasma beta range from $0.016$ to $0.08$. 
 The domain size is $L\times L \times L$ = $102.4 d_{\rm i}\times 102.4 d_{\rm i} \times 102.4 d_{\rm i}$. The resolution of the simulations is $N_x \times N_y \times N_z = 2048\times 1 \times 2048$ for $m_i/m_e =25 $ and $4096\times 1 \times 4096$ for $m_i/m_e =100$. 
The number of macroparticles for each species per cell is 400. Periodic boundary conditions are used.
Initially, the magnetic fluctuation in all simulations contains the total energy of $5.4\times 10^5\, b^2_0 d^3_i/\mu_0$.
Simulations were run until the magnetic free energy dropped to $\sim 12\%$ of the total energy from the initial magnetic fluctuation. All our reported results are for $m_i/m_e$ = 25 except where noted.

\section{\label{result}Results}

\begin{figure*}[ht]
  \begin{center}
    \includegraphics[width=1.0\textwidth]{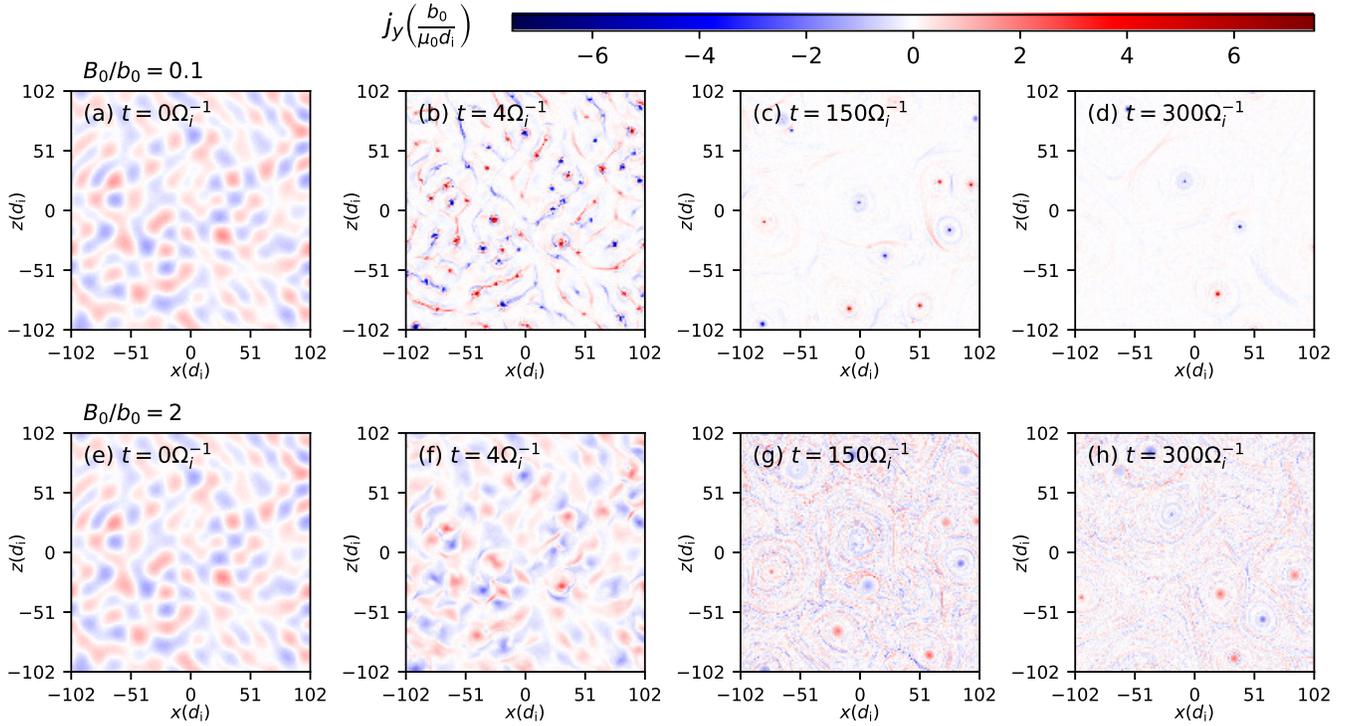}
    \caption{\small The $y$-component of the current density at $\Omega_i t=0$, $4$, $150$, and $300$ from the simulations with $m_i/m_e = 25$ and $B_0/b_0=0.1 $ (top) and $B_0/b_0=2$ (bottom).}
\label{fig:jy}
  \end{center}
\end{figure*}

\subsection{Plasma Dynamics}

\begin{figure}[ht]
  \begin{center}
    \includegraphics[width=0.9\textwidth]{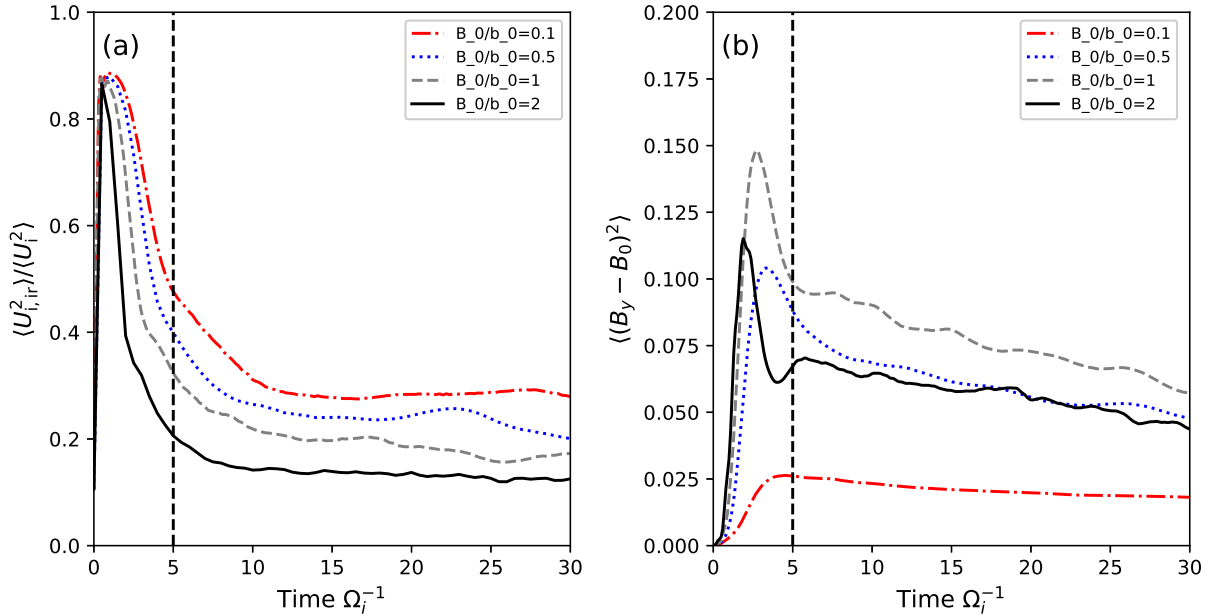}
    \caption{\small (a) Compressible fraction of the ion flow energy $\langle U^2_{\rm i,ir}/U^2_{\rm i}\rangle$ as a function of time. (b) Spatial average of the square of the magnetic fluctuation in the $y$ direction ,i.e., the free energy of the guide magnetic field. Dashed line indicates time $\tau$ between the island-forming and decay phases.}
\label{fig:div}
  \end{center}
\end{figure}

Figure~\ref{fig:jy} shows the $y$-component of the current density, $j_y$, at 
$\Omega_i t=0$, $4$, $150$ and $300$ from the simulations with $B_0/b_0=0.1$ and $2$.
The initial plasma configuration is unstable since the forces are extremely imbalanced. 
At this early time, the plasma flow is highly compressible as current structures contract to form magnetic islands.
Figures~\ref{fig:div}(a) shows the compressible fraction of the ion flow energy $\langle U^2_{\rm i,ir}
\rangle/\langle U^2_{\rm i}\rangle$, which is the ratio of the kinetic energy of the compressible (irrotational) ion flow $U_{\rm i,ir}$ to the total kinetic energy of the ion flow $U_{\rm i}$.
This ratio quickly rises at the beginning and reaches a maximum value greater than 0.8 after a few ion cyclotron times in all simulations. Later, the ratio declines and remains fairly stationary after $\Omega_{\rm i}\tau=5$ (dashed line).
The current contraction also builds free energy in $B_y$ as shown in Figure~\ref{fig:div}(b). From the simulations, the magnitude of $B_y-B_0$ is high at the centers of the magnetic islands.
After a few ion cyclotron times, the current within each flux rope contracts, reducing energy in the magnetic field. 
Multiple major magnetic islands become obvious after the contraction of plasma currents inside each island as shown in  Figure~\ref{fig:jy} at $\Omega_{\rm i}t=4$, $150$, and $300$.

\add{The conditions at  $t\lesssim\tau$, the island-forming phase, involve imbalanced forces that reflect the initial condition and are unlikely to appear in nature.} 
At $t\gtrsim\tau$ \add {, the current-merging or decaying phase}, the forces are only weakly imbalanced, and the root-mean-square of current density reaches its maximum value at $\Omega_{\rm i}t=25$. 
\add{In this phase, multiple magnetic islands merge as the currents sharing the same direction attract one another. Before the merging of two islands is complete, a current sheet emerges between them and becomes a site for magnetic reconnection accompanied with a number of small islands.} 
This phase is more applicable to natural situations in space plasmas.
For stronger $B_0$, there are more of these tiny islands; see the plots of $j_y$ at $\Omega_i t=150$ and $300$ in Figure~\ref{fig:jy}. 
These emerging plasmoids from the reconnection have a short lifetime as they are destroyed by smaller-scale magnetic reconnection  after colliding with the surrounding plasma. \add {As the current merging continues, there are fewer sites for magnetic reconnection and associated plasma activity}

\subsection{Energy Conversion Profiles}
\subsubsection{Overview}
\begin{figure}[ht]
  \begin{center}
    \includegraphics[width=0.9\textwidth]{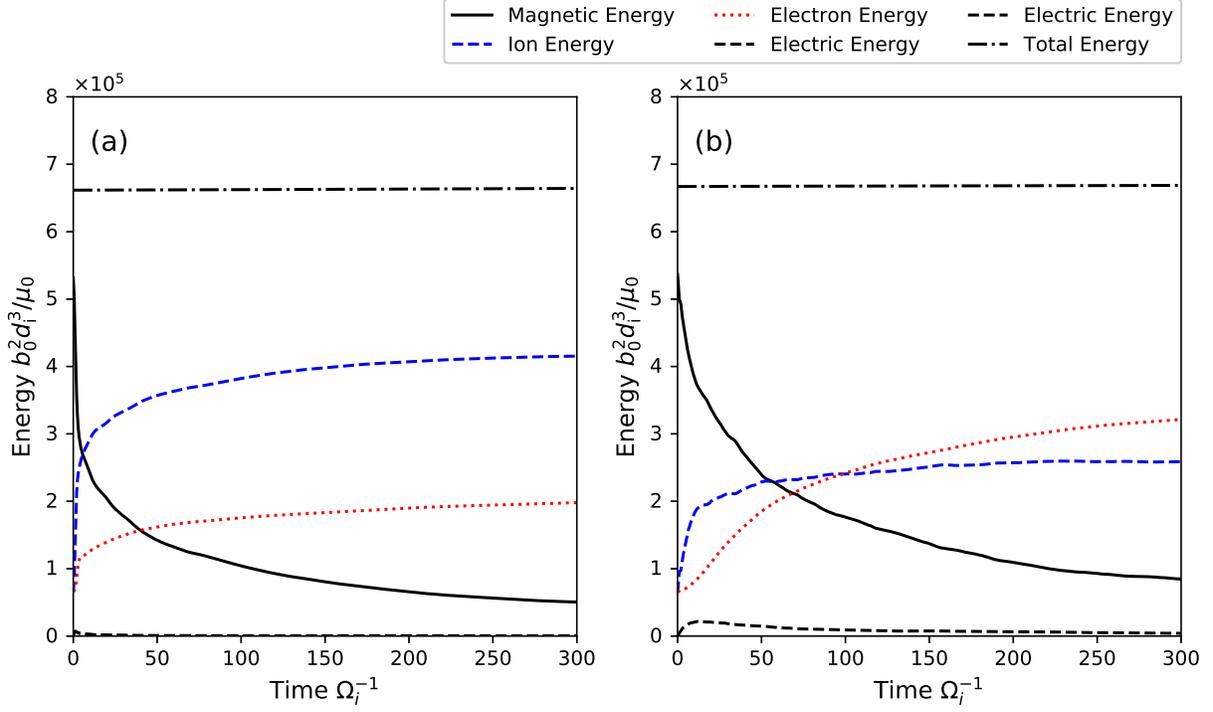}
    \caption{\small Time series of the total magnetic fluctuation energy, total electric energy, total ion energy, total electron energy, \add{ total transferable energy} and (a) $B_0/b_0=0.1$ and (b) $B_0/b_0=2$.}
\label{fig:energies}
  \end{center}
\end{figure}

Figure~\ref{fig:energies} shows the magnetic fluctuation energy, electric field energy, total ion energy, total electron energy, and \add{total transferable energy} as time series from the simulations with $B_0/b_0=0.1$ and $2$. 
The electric energy is tiny and nearly constant, so it does not serve as a significant source or sink of energy, and refering to Figure~\ref{fig:conversion},  ${\pp J}\cdot{\pp E}$ can accurately represent both electric and magnetic energy conversion rates per unit volume.   
When $B_0$ changes, the overall amount of energy gained by ions and electrons also changes. 
Electrons eventually gain more energy for higher $B_0$~\citep{gary16,hughes17}.
In our 2.5D simulations, all quantities depend on $x$ and $z$ while $y$ is an ignorable coordinate. By applying Faraday's law to our scenarios, only the $y$ component of the electric field can modify the in-plane magnetic field while the in-plane electric field can only modify $B_y$. Since the initial magnetic fluctuation is in the $x-z$ plane, the overall conversion of energy from the magnetic field is well represented by the integral of $J_yE_y$ over space and time. This integration includes magnetic energy lost during the island-forming and the decaying phases.

At $t\lesssim \tau$, energy in $B_x$ and $B_z$ drops via $E_y$ but the energy in $B_y$ increases. During the island-forming phase, $\langle J_yE_y\rangle$ is higher than $\langle {\pp J}\cdot{\pp E}\rangle$ because part of energy from $\langle J_yE_y\rangle$ is used for building strong free energy $\langle(B_y-B_0)^2\rangle$ at centers of the magnetic islands via the in-plane solenoidal electric field ${\pp E}_{{\rm so},xz}$.
Therefore, $\langle J_yE_y\rangle$ must be less than $\langle {\pp J}\cdot{\pp E}\rangle$ during the decaying phase. 

\begin{figure}[ht]
  \begin{center}
    \includegraphics[width=0.9\textwidth]{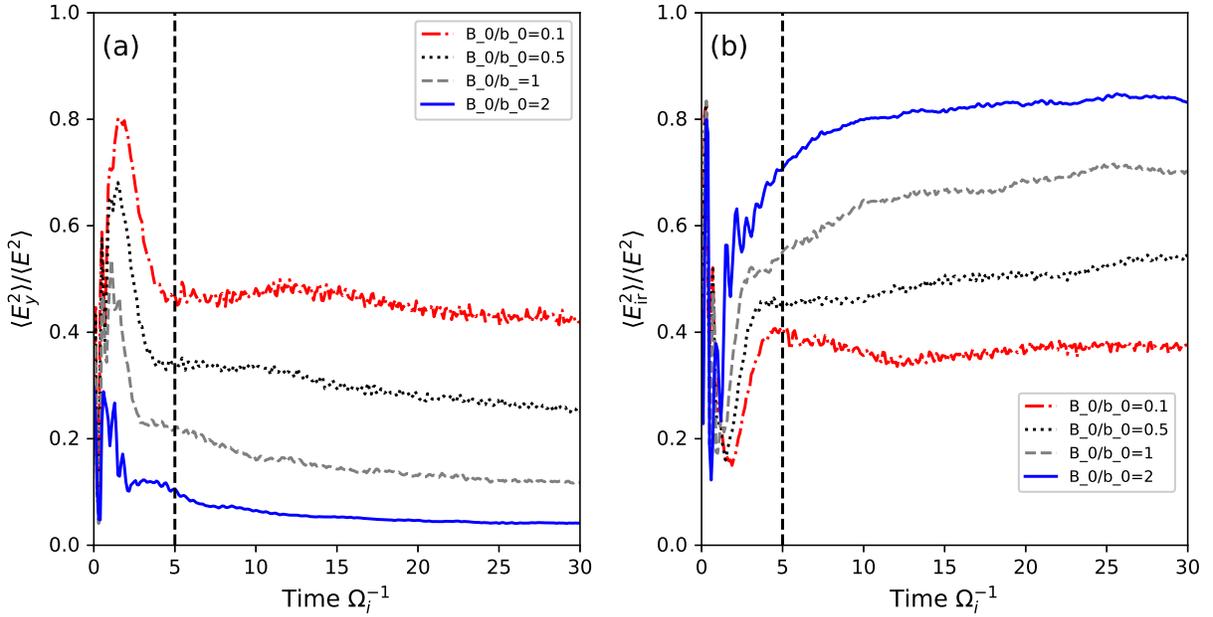}
    \caption{\small Fraction of electric energy in the (a) $y-$component and (b) irrotational component as a function of time. Dashed line indicates time $\tau$ between island-forming and decaying phases.}
\label{fig:e_energy}
  \end{center}
\end{figure}
Since the island-forming phase and decaying phase have distinct mechanisms of energy conversion, these can be analyzed separately. As the decaying phase is more applicable to astrophysical situations, we focus more on the energy conversion during this later phase of the simulation. We examine the mechanism that takes energy out of $B_y$ while $E_y$ continues to take energy out of the in-plane components of the magnetic field.

Figures~\ref{fig:e_energy}(a) shows the ratio of the energy in $E_y$ to the total electric field energy at times up to $30 \Omega^{-1}_i$. 
As $B_0$ is increased, $E_y$ provides a lower energy fraction at $t\gtrsim \tau$. 
For $B_0/b_0=2$, the energy in $E_y$ is around $5\%$ of the total electric field energy at $t\gtrsim \tau$. 
Meanwhile, the energy in ${\pp E}_{\rm ir}$ can be as high as $80\%$ of the total electric field energy for $B_0/b_0=2$ at $t\gtrsim \tau$, as shown in Figure~\ref{fig:e_energy}(b). 
Since ${\pp E}_{\rm ir}$ provides a significant fraction of the electric energy, which even dominates at large $B_0$, a large portion of ${\pp J}\cdot{\pp E}$ represents energy exchange between ions and electrons without directly involving the conversion of magnetic energy. The analysis of how ${\pp J}\cdot{\pp E}$ take energy out of the magnetic field must consider the large contribution from ${\pp E}_{\rm ir}$ into account.

\subsubsection{Island-forming Phase}
\begin{figure*}[ht]
  \begin{center}
    \includegraphics[width=0.9\textwidth]{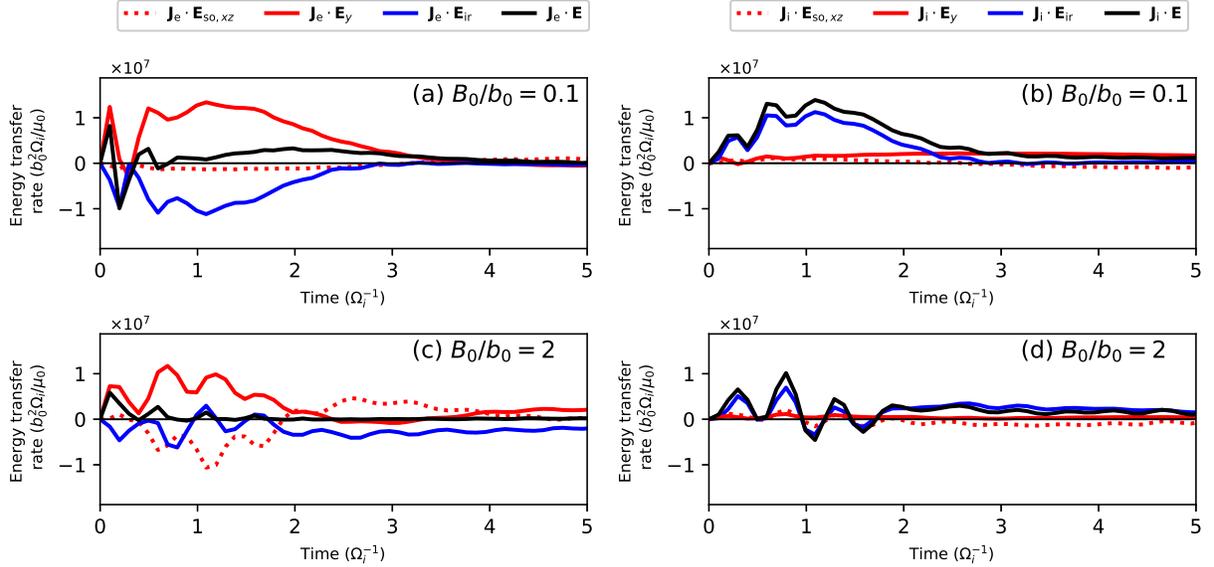}
    \caption{\small Panels (a) and (b) show the rates of energy  transfer to ions and electrons, respectively, for various electric field components with $B_0/b_0=0.1$ for $t<\tau$. Panels (c) and (d) show the same rates to ions and electrons, respectively, with $B_0/b_0=2$.
    }
\label{fig:rates_early}
  \end{center}
\end{figure*}

First consider $\langle {\pp J}\cdot{\pp E}\rangle$ at $t<\tau$.
We separately consider the energy transfer to ions and electrons as well as the contributions from ${\pp E}_{\rm ir}$, ${\pp E}_{{\rm so},xz}$ and $E_y$ to $\langle {\pp J}\cdot{\pp E}\rangle$ (Figure~\ref{fig:rates_early}). (Note that ${\bf E}_y$ is purely solenoidal.) At $t\lesssim\tau$, simulations with $B_0/b_0=0.1$ and $2$ both show that electrons are mainly responsible for taking energy out of the in-plane magnetic field via ${\bf E}_y$. Electrons then transfer most of this energy to ions via ${\pp E}_{\rm ir}$ (recall that ${\pp E}_{\rm ir}$ is associated with charge separation) and ions gain more energy than electrons in this phase.
The profiles of the energy transfer rates to electrons and ions via ${\pp E}_{\rm ir}$ are very similar but with opposite signs. 
This implies that the particle interactions with ${\pp E}_{\rm ir}$ {mostly do not store or extract electric} energy in ${\pp E}_{\rm ir}$ but only cause the energy exchange between particles of opposite charges.  

For simulations with $m_i/m_e=25$, and $B_0/b_0=0.1$, $0.5$, $1$ and $2$, the total energy $W_{25,1}$ gained by particles during the $1^{st}$ (island-forming) phase from the beginning to $t=\tau$ was $2.5\times 10^5$, $1.9\times 10^5$, $1.3\times 10^5$ and $1.0\times 10^5$, respectively in units of $b^2_0 d^3_{\rm i}/\mu_0$.
Less energy is transferred to particles during this phase when $B_0$ is higher. Table~\ref{tab:erate_25} shows the fraction of the total electromagnetic energy transfer to particles that go to ions and electrons via 
${\pp E}_{\rm ir}$, ${\pp E}_{{\rm so},xz}$, $E_y$ and the total eletric field ${\pp E}$ 
from the beginning to $t=\tau$, for each simulation.
\begin{table}
\caption{\label{tab:erate_25}
The fractions of the total electromagnetic energy transfer to particle energy that go to ions and electrons during the island-forming phase $(t<\tau)$ via interactions with ${\pp E}_{\rm so,xz}$, $E_y$, ${\pp E}_{\rm ir}$, and ${\pp E}$.}

\begin{ruledtabular}
\begin{tabular}{cccccc}
$B_0/b_0$ & Species &${\pp J}_{\rm s}\cdot {\pp E}_{{\rm so}, xz}$ & ${\pp J}_{\rm s}\cdot {\pp E}_y$ & ${\pp J}_{\rm s} \cdot {\pp E}_{\rm ir}$ & ${\pp J}_{\rm s} \cdot {\pp E}$ \\
\hline
\multirow{2}{*}{0.1} & Ion & -0.01 & 0.28 &0.52 & 0.79 \\
& Electron & -0.05 &  0.78 & -0.52 & 0.21\\
\hline
\multirow{2}{*}{0.5} & Ion & -0.08 &0.35 &0.55 & 0.82 \\
& Electron & -0.17 &0.92 &-0.57 & 0.18 \\
\hline
\multirow{2}{*}{1} & Ion & -0.18 &0.30 & 0.76 & 0.88 \\
& Electron & -0.26 & 1.17 &-0.79 & 0.12 \\
\hline
\multirow{2}{*}{2} & Ion & -0.24 &0.19 & 0.88 & 0.83 \\
& Electron & -0.13 & 1.28 &-0.98 & 0.17 \\
\end{tabular}
\end{ruledtabular}

\end{table}

In all simulations, electrons take energy from the in-plane components of the magnetic field and transfer a large amount of energy to ions via ${\pp E}_{\rm ir}$. Both ions and electrons lose some energy to form strong $|B_y-B_0|$ at the centers of magnetic islands. As $B_0/b_0$ is increased up to 1, more energy is transferred to $B_y$ and the energy transfer via ${\pp E}_{\rm ir}$ is also stronger. At $t=\tau$, ions have gained more energy than electrons even though electrons are the main species interacting with $E_y$ and reducing the magnitude of the in-plane magnetic field. 

\subsubsection{Decaying Phase}
At $t > \tau$, the current contraction is less active.
Magnetic energy drops due to the merging of magnetic islands.
We carry on the analysis of energy conversion by considering the contributions from ions and electrons, again separating effects of ${\pp E}_{\rm ir}$ from ${\pp E}_{\rm so}$. 
In this phase, we found that separating the solenoidal field into ${\bf E}_y$ and ${\pp E}_{{\rm so},xz}$ is not the most effective way to understand the energy conversion. In this phase, energy conversion often takes place where both $B_y$ and the in-plane magnetic field lose energy together. We found that separating the solenoidal electric field into the components parallel ${\pp E}_{\rm so,\parallel}$ and perpendicular ${\pp E}_{\rm so,\perp}$ to the local magnetic field is more suitable for a strong guide field.
\footnote{We have considered the relevance of components parallel and perpendicular to {\bf B} vs.\ out-of-plane ($y$) and in-plane ($x$-$z$) components for the decaying phase.  We found that resolving components parallel and perpendicular to {\bf B} provides a clearer physical distinction for particle acceleration (direct parallel acceleration vs.\ perpendicular drift motion) and also a clearer distinction among components of the pressure tensor.}

\begin{figure*}[ht]
  \begin{center}
    \includegraphics[width=1.0\textwidth]{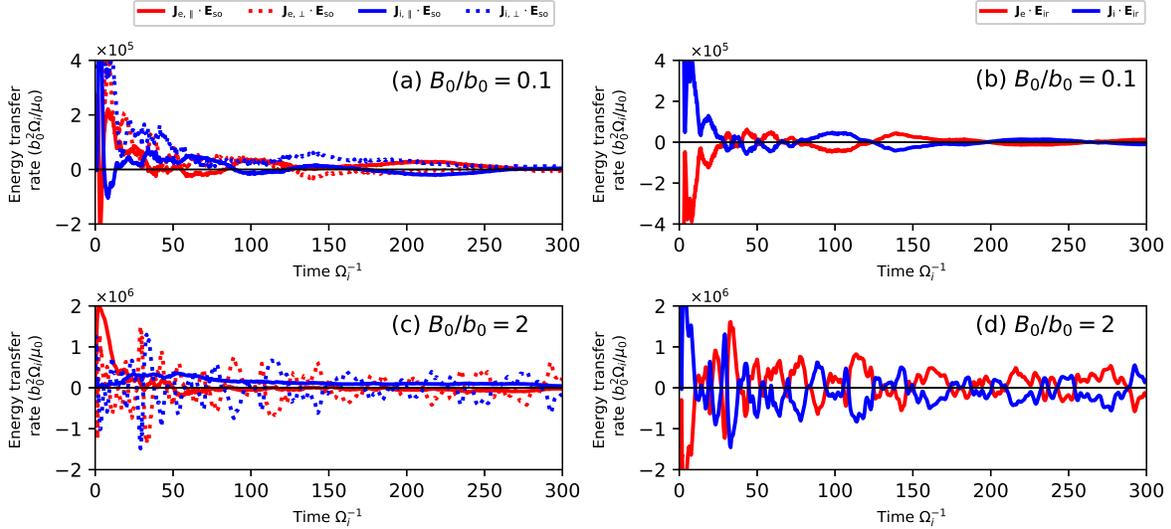}
    \caption{\small The rates of energy conversion to ions and electrons by interacting with (a) components of ${\pp E}_{\rm so}$ and (b) ${\pp E}_{\rm ir}$ from the simulations with $B_0/b_0=0.1$ while panels (c) and (d) shows the energy conversion rates to ions and electrons from interaction with components of ${\pp E}_{\rm so}$ and ${\pp E}_{\rm ir}$, respectively, with $B_0/b_0=2$. 
    }
\label{fig:rates_later}
  \end{center}
\end{figure*}

Figures~\ref{fig:rates_later}(a) and ~\ref{fig:rates_later}(c) show the energy conversion rates to electrons and ions via ${\pp E}_{\rm so}$ for $B_0/b_0=0.1$ and $2$, respectively, at $t>\tau$. For $B_0/b_0=0.1$, $\langle {\pp J}_{i,\perp}\cdot{\pp E}_{\rm so}\rangle$ is positive and dominates other rates.
For $B_0/b_0=2$, $\langle {\pp J}_{e,\perp}\cdot{\pp E}_{\rm so}\rangle$ and $\langle {\pp J}_{i,\perp}\cdot{\pp E}_{\rm so}\rangle$ fluctuate with large magnitudes, share similar shapes and seems to have the opposite signs. 
On the other hand, the rates $\langle {\pp J}_{e,\parallel}\cdot{\pp E}_{\rm so}\rangle$ and $\langle {\pp J}_{i,\parallel}\cdot{\pp E}_{\rm so}\rangle$  remain mostly positive with smaller magnitudes.
The rates from ${\pp E}_{\rm ir}$ are shown in Figure~\ref{fig:rates_later}(b) for $B_0/b_0=0.1$ and Figure~\ref{fig:rates_later}(d) for $B_0/b_0=2$. These rates fluctuate with stronger magnitude when $B_0$ is higher. The rates for ions and electrons are nearly equal and opposite almost all the time, representing energy transfer among particle species with little net conversion of electromagnetic energy to particle motion.

Let $W_{25,2}$ be the total energy gained by particles in the $2^{nd}$ (decaying) phase after $t=\tau$ for each simulation with $m_i/m_e=25$. The values are $2.3\times 10^5$, $2.8\times 10^5$, $3.4\times 10^5$ and $3.7\times 10^5$ in units of $b^2_0 d^3_{\rm i}/\mu_0$ for $B_0/b_0=0.1$, $0.5$, $1$ and $2$, respectively.
Table~\ref{tab:erate_25_2} shows the fractions of the total energy gain that go to ions and electrons via  ${\pp E}_{\rm so,\parallel}$, ${\pp E}_{\rm so,\perp}$, ${\pp E}_{\rm ir}$ and the total electric field ${\pp E}$.
The values in the table are fractions of $W_{25,2}$. 
For $B_0/b_0=0.1$, the most effective motion to transfer energy via ${\pp E}_{\rm so}$ is the perpendicular motion of ions. This motion takes $0.51W_{25,2}$ of the converted energy while the parallel and the perpendicular motions of electrons take $\sim 0.20W_{25,2}$ each. 
The interaction between ions and electrons via ${\pp E}_{\rm ir}$ is weak. The total amounts of energy gained by ions and electrons for $t>\tau$ are $0.65W_{25,2}$ and $0.35W_{25,2}$, respectively. 
For small $B_0$ at $t>\tau$, the energy transferred via ${\bf E}_y$ still dominates and $92\%$ of electromagnetic energy conversion is via ${\bf E}_y$. Therefore, separating the rates from ${\pp E}_{\rm so,\parallel}$ and ${\pp E}_{\rm so,\perp}$ is not particularly useful.
As $B_0$ increases, the parallel motion of ions becomes the most effective motion to take energy from ${\pp E}_{\rm so}$. The particle interaction through ${\pp E}_{\rm ir}$ is also stronger as ${\pp E}_{\rm ir}$ is stronger for higher $B_0$. For $B_0/b_0=2$, the parallel motion of ions takes $0.91W_{25,2}$ from ${\bf E}_{\rm so}$ while ${\bf E}_{\rm ir}$ takes $0.67W_{25,2}$ from ions and gives $0.68W_{25,2}$ to electrons. For $t>\tau$, the energy gained by electrons is $0.73W_{25,2}$ while ions gain only $0.27W_{25,2}$. As $B_0$  increases, the energy conversion by the parallel electric field becomes stronger. For $B_0/b_0=2$, ${\pp E}_{\rm so,\parallel}$ is responsible for almost $100\%$ of $W_{25,2}$ while $E_y$ takes $88\%$ of $W_{25,2}$ out of the magnetic field. The dominance of ${\pp E}_{\rm so,\parallel}$ implies that energy in $B_y$ and the in-plane magnetic field are likely drained at the same locations via ${\pp E}_{\rm so,\parallel}$.

For comparison, we calculate the energy conversion via ${\pp E}_{\rm so,\parallel}$ and ${\pp E}_{\rm so,\perp}$ in Table \ref{tab:erate_25_before} for $t<\tau$.
For $B_0/b_0=0.1$, the magnetic field lies mostly in the $x$-$z$ plane initially and the perpendicular direction is nearly along the $y$ direction. Therefore, in the first phase, the sum of ${\pp J}_{\perp}\cdot {\pp E}_{\rm so,\perp}$ from both species, with $E_y$ as the main component in ${\pp E}_{\rm so,\perp}$, is responsible for $95\%$ of the energy conversion. For the case of $B_0/b_0=2$, the initial magnetic field tends to be more aligned with the $y$ direction.
As $E_y$ is the main component of ${\pp E}_{\rm so,\parallel}$, the sum of ${\pp J}_{\parallel}\cdot {\pp E}_{\rm so,\parallel}$ from both species is responsible for $89\%$ of the energy conversion. Thus these results are consistent with our conclusion from the previous section that $E_y$ plays the dominant role in energy conversion for $t<\tau$ (see Table~\ref{tab:erate_25}).

The energy transfer processes in the phases $t<\tau$ and $t>\tau$ are very different. Electrons are the main species to draw energy from the magnetic field in the early phase but lose most of the energy to ions via the charge separation and ${\bf E}_{\rm ir}$. In the later phase, ions are more responsible for draining energy from the magnetic field. For $B_0/b_0=0.5$, $1$ and  $2$, these ions lose energy to electrons via the charge separation and ${\bf E}_{\rm ir}$.
The interaction between ions and electrons via ${\bf E}_{\rm ir}$ is stronger as $B_0$ increases.
\add{ Our analysis is insensitive to choices of the boundary time from $\Omega_{\rm i}\tau=5$ to $10$. By using $\Omega_{\rm i}\tau=10$, the results change numerically but remain qualitatively similar. The results for $m_i/m_e=100$ are also qualitatively similar to those for $m_i/m_e=25$.}

\begin{table}
\caption{\label{tab:erate_25_2}
The fraction of the total electromagnetic energy transfer to particle energy that go to ions and electrons via interactions with with ${\pp E}_{\rm so,\parallel}$, ${\pp E}_{\rm so,\perp}$, ${\pp E}_{\rm ir}$, and ${\pp E}$ during the decaying phase ($t>\tau$).}
\begin{ruledtabular}
\begin{tabular}{cccccc}
$B_0/b_0$ & Species &${\pp J}_{\rm s,\parallel}\cdot {\pp E}_{\rm so,\parallel}$ & ${\pp J}_{\rm s,\perp}\cdot {\pp E}_{\rm so,\perp}$ & ${\pp J}_{\rm s} \cdot {\pp E}_{\rm ir}$ & ${\pp J}_{\rm s} \cdot {\pp E}$ \\
\hline
\multirow{2}{*}{0.1} & Ions & 0.04 & 0.51 & 0.10 & 0.65 \\
& Electrons & 0.20 &  0.23 &-0.08 & 0.35\\
\hline
\multirow{2}{*}{0.5} & Ions & 0.63 &0.02 & -0.07 & 0.58 \\
& Electrons & 0.16 & 0.18 &0.08 & 0.42 \\
\hline
\multirow{2}{*}{1} & Ions & 1.21 & -0.01 & -0.68 & 0.52 \\
& Electrons & -0.27 & 0.05 & 0.70 & 0.48 \\
\hline
\multirow{2}{*}{2} & Ions & 0.91 & 0.03 & -0.67 & 0.27 \\
& Electrons & 0.10  &-0.05 &0.68 & 0.73 \\
\end{tabular}
\end{ruledtabular}

\end{table}

\begin{table}
\caption{\label{tab:erate_25_before}
The fraction of the total electromagnetic energy transfer to particle energy that go to ions and electrons via interactions with ${\pp E}_{\rm so,\parallel}$ and ${\pp E}_{\rm so,\perp}$ for the island-forming phase ($t<\tau$).}
\begin{ruledtabular}
\begin{tabular}{cccc}
$B_0/b_0$ & Species &${\pp J}_{\rm s,\parallel}\cdot {\pp E}_{\rm so,\parallel}$ & ${\pp J}_{\rm s,\perp}\cdot {\pp E}_{\rm so,\perp}$ \\
\hline
\multirow{2}{*}{0.1} & Ions & 0.06 & 0.21 \\
& Electrons & $\approx 0$ &  0.73 \\
\hline
\multirow{2}{*}{0.5} & Ions & 0.01 &0.26 \\
& Electrons & 0.23 &0.52 \\
\hline
\multirow{2}{*}{1} & Ions & -0.02 &0.14 \\
& Electrons & 0.48 & 0.43 \\
\hline
\multirow{2}{*}{2} & Ions & 0.03&-0.08 \\
& Electrons & 0.86  &0.29 \\
\end{tabular}
\end{ruledtabular}

\end{table}

\subsection{Indicators of Energy Conversion Locations}
From the previous section, the overall rate of energy transfer to particles in the decaying phase is approximately equal to the integration of $J_\parallel E_{\rm so, \parallel}$ over $t>\tau$. 
In all the simulations, $\nabla\times{\pp B} \approx \mu_0{\pp J}$ is valid at most locations so we can write
\begin{equation}
J_\parallel E_{\rm so, \parallel} \approx 
  \frac{1}{\mu_0} {\pp E}_{\rm so, \parallel}\cdot\nabla\times{\pp B}  =
  \frac{1}{\mu_0}{\pp B}\cdot\nabla\times{\pp E}_{\rm so, \parallel},   
\label{eq:rates_b2}
\end{equation}
noting that ${\pp E}_{\rm so,\parallel}$ is parallel to ${\pp B}$, so there is no Poynting vector associated with ${\pp E}_{\rm so,\parallel}$. 
The last term is the rate of change of the local magnetic energy density due to ${\pp E}_{\rm so, \parallel}$ from Faraday's law.
Therefore, the energy transferred via the parallel motion is local and the magnetic energy quite instantly becomes the kinetic energy of particles where $\nabla\times{\pp B}\approx \mu_0{\pp J}$.

\begin{figure}[ht]
  \begin{center}
    \includegraphics[width=0.9\textwidth]{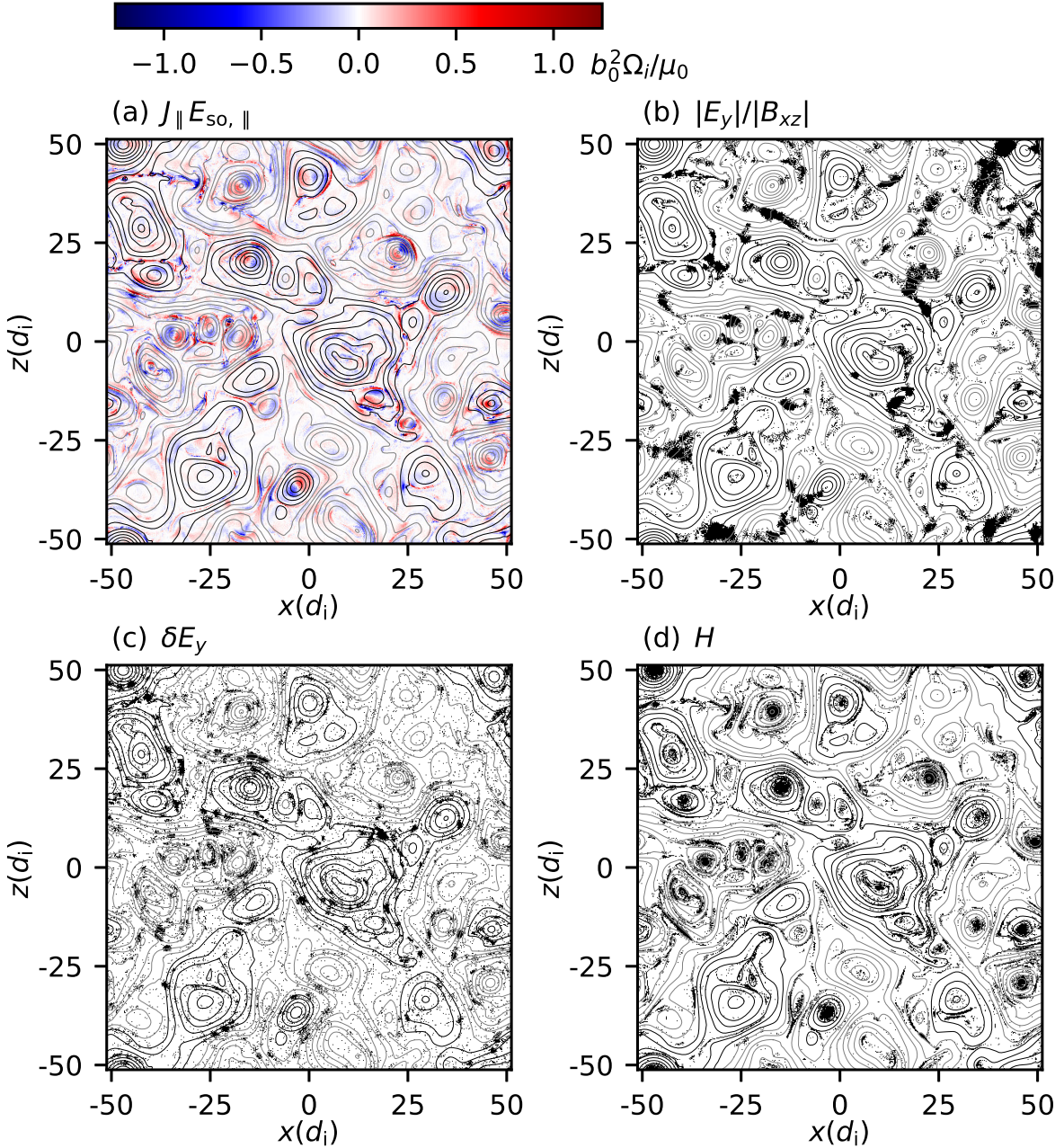}
    \caption{\small (a) 2D color plot of $J_\parallel E_{\rm so,\parallel}$ at $t=22\Omega^{-1}_i$.
     (b)-(d) Spatial regions (black) where the indicator is (b) $|E_y|/|B_{xz}|$ less than its 10th percentile, or (c) $\delta E_y$ or (d) $H$ greater than its 90th percentile.
    }
\label{fig:ej_locations}
  \end{center}
\end{figure}

Figure~\ref{fig:ej_locations}(a) shows the color plot of $J_\parallel E_{\rm so, \parallel}$ over the entire domain at $t=22\Omega^{-1}_i$ from the simulations with $B_0/b_0=2$.
Contours of vector potential $A\hat{y}$ are shown as black $(A<0)$ and grey $(A>0)$.  In 2D simulations, these curves are also the projection of the magnetic field lines on the $x$-$z$ plane.
The magnetic islands are surrounded by circular closed curves while the current sheets are located at boundaries between magnetic islands that share the same rotation direction. The rate $J_\parallel E_{\rm so, \parallel}$ fluctuates strongly and is highly intermittent in space. 
The energy is transferred strongly both ways, either from fields to particles ($J_{\rm parallel} E_{\rm so,\parallel}>0$, red color) or from particles to fields ($J_{\rm parallel} E_{\rm so,\parallel}<0$, blue color).
From the previous sections we have learned that the spatial average of $J_\parallel E_{so,\parallel}$ is positive, indicating a net conversion of energy from fields to particle kinetic energy.
We can see two types of structures with strong (positive or negative) values of $J_\parallel E_{so,\parallel}$: 
1) Near the centers of magnetic islands, there are bipolar structures, with positive and negative $J_\parallel E_{so,\parallel}$.
We have confirmed that the positive and negative values nearly cancel, and these regions make a tiny contribution to the net energy conversion rate $\langle J_\parallel E_{so,\parallel}\rangle$.
2) Near the edges of magnetic islands, there are elongated regions with strong $J_\parallel E_{so,\parallel}$, including regions of magnetic reconnection.  
The positive values of $J_\parallel E_{so,\parallel}$ dominate and these regions account for the overall positive energy conversion.Therefore, it would be useful to find quantities that indicate these regions near the edges of magnetic islands, especially if they provide some insight into why the energy conversion mostly occurs in these specific locations.

In this section, we examine various quantities as possible indicators of energy conversion locations, and specifically of regions with strong and positive $J_\parallel E_{\rm so, \parallel}$ on average. 
 We test these indicators using the following steps.
 For any continuous indicator $X$, we identify the median $X_m$ of $X$ values at all grid points over the spatial domain as a time series (for $t>\tau$). We integrate the total energy $\epsilon_{X,h}$ transferred to particle kinetic energy by $J_\parallel E_{\rm so,\parallel}$ from the region with $X>X_m$ over the time domain of interest and compare $\epsilon_{X,h}$ with the total energy $\epsilon_X$ from $J_\parallel E_{\rm so,\parallel}$ over the spatial and time domains. We then calculate
 a measure of predictive power,
 $S_X=2|\epsilon_{X,h}/\epsilon_X-0.5|$. 
 Because a monotonic transformation of $X$ does not change its percentiles, with this construction, $S_x$ is invariant upon any monotonic transformation of the indicator $X$, and $0\leq S_X\leq 1$.
 If $S_X$ is close to $1$, then $X$ is a good indicator of energy conversion locations.
 
We have {tested many} indicators using the data from simulations with $B_0/b_0=2$. 
Table~\ref{tab:indicators} shows a list of indicators and their predictive power regarding energy transferred via $J_\parallel E_{\rm so,\parallel}$ for all times $t>\tau$. We include $|E_{\rm so,\parallel}|$ and $|J_\parallel|$, which multiply to make $|J_\parallel||E_{\rm so,\parallel}|$, only for comparison.
We found three good indicators in addition to $|E_{\rm so,\parallel}|$ and $|J_\parallel|$ themselves.

\begin{table}
\caption{\label{tab:indicators}
Indicators $X$ and a measure of their predictive power, $S_X$, based on the fraction of the energy conversion $J_\parallel E_{\rm so,\parallel}$ that occurs in the region with $X$ greater than its median $X_m$. $|J_\parallel|$ \& $|E_{\rm so,\parallel}|$ themselves, indicated in italics, are included for comparison purposes.}
\begin{ruledtabular}
\begin{tabular}{lcc}
Indicator & Formula & $S_X$\\
\hline

{\it Magnitude of parallel current} & $|{\pp J}_\parallel|$ & 0.9 \\ 

Magnitude of ratio of out-of-plane electric field to in-plane magnetic field &  $|E_y|/|B_{xz}|$
 &
0.82 \\ 

{\it Magnitude of parallel component of solenoidal electric field} &$|{\pp E}_{\rm so,\parallel}|$ & 0.8 \\ 

Magnitude of $y$-component of non-ideal electric field & $\delta E_y=|{\pp E}_{y}+({\pp U}\times{\pp B})_{y,\rm so}|$ &
0.8 \\ 

Estimate of current helicity &  $H=|{\pp B}\cdot\nabla\times{\pp B}/\mu_0|$ & 0.76 \\  

Magnitude of in-plane component of magnetic fluctuation & $|B^2_x+B^2_z|$ & 0.46 \\ 

Magnitude of vorticity & $|\nabla\times {\pp U}|$ &
0.46 \\ 

Magnitude of solenoidal component of non-ideal electric field & $|{\pp E}_{\rm so}+({\pp U}\times{\pp B})_{\rm so}|$ &
0.36 \\  

Magnitude of non-ideal electric field & $|{\pp E}+{\pp U}\times{\pp B}|$ &
0.34 \\  

Divergence of bulk velocity & $\nabla\cdot {\pp U}$ &
0.26 \\  

Magnitude of cross helicity  & $|{\pp U}\cdot({\pp B}-B_0\hat{y})|$ & 0.1 \\ 

Magnitude of $y$-component of magnetic fluctuation & $|B_y-B_0|$ & 0.06 \\ 

Magnitude of curvature of magnetic field & $|\hat{B}\cdot\nabla \hat{B}|$ & 0.06 \\ 

Magnitude of magnetic helicity & $|{\pp A}\cdot{\pp B}|$ & 0.06 \\ 

Magnitude of magnetic field & $|B|$ & 0 \\ 

\end{tabular}
\end{ruledtabular}

\end{table}
The best such indicator is $|E_y|/|B_{xz}|$, which is based on an indicator suggested by~\cite{lapenta2021}.
That work actually examined the formal transformation velocity ${\bf v}_L=c^2{\bf B}\times{\bf E}/E^2$ such that a Lorentz transformation would nullify the two components of the magnetic field in the plane perpendicular to the electric field, as an indicator of reconnection regions in 3D space.
The magnitude of ${\bf v}_L$ is almost always superluminal, with $v_L<c$ only in the vicinity of reconnection sites.
For our configuration, with a guide magnetic field component $B_y$, reconnection sites should have ${\bf B}_{xz}\approx0$ and we can simply use $v_L=c^2|B_{xz}|/|E_y|$.  
As a further simplification, we divide $v_L$ by $c^2$ and invert to obtain $|E_y|/|B_{xz}|$, which has units of velocity and is usually close to zero, but is large near null points of ${\bf B}_{xz}$ that have non-zero $E_y$, i.e., near reconnection regions.

 We found that $S_{|E_y|/|B_{xz}|}$ is 0.82,i.e., $91\%$ of the energy transferred via $J_\parallel E_{\rm so, \parallel}$ occurs in  regions with $|E_y|/|B_{xz}|$ higher than its median value, which are typically near reconnection sites. This indicator has higher predictive power for $J_\parallel E_{\rm so, \parallel}$ than $|E_{\rm so, \parallel}|$ itself.

The second best indicator is the magnitude of the $y-$component of the non-ideal electric field, $\delta E_y =\hat{y}\cdot({\pp E}+{\pp U}\times {\pp B})$, where ${\pp U}$ is the plasma bulk velocity. 
At locations with a high magnitude of ${\pp E}+{\pp U}\times {\pp B}$, the frozen-in approximation is no longer valid. However, we find that the magnitude of ${\pp E}+{\pp U}\times {\pp B}$ is not as good  indicator as $\delta E_y$. We found that $S_{\delta E_y}$ is 0.8,  implying that $90\%$ of the energy transferred via $J_\parallel E_{\rm so, \parallel}$ occurs in regions with $\delta E_y$ above its median value. This indicator has a predictive power for $J_\parallel E_{\rm so,\parallel}$ equal to $|E_{\pm so,\parallel}|$ itself. 
 This implies that strong energy conversion often takes place where the frozen-in approximation fails which also typically occurs near reconnection sites.

The third best indicator is an estimate $H=|{\pp B}\cdot\nabla\times{\pp B}|/\mu_0$ of the current helicity $|{\bf J}\cdot{\bf B}|$. We can write
\begin{equation}
J_\parallel E_{\rm so, \parallel}=\frac{E_{\rm so,\parallel}|{\bf J}\cdot{\bf B}| }{B},
\label{eq:rates_b2}
\end{equation}
so the current helicity is directly linked with
$J_\parallel E_{\rm so, \parallel}$.
We use $H=|{\pp B}\cdot\nabla\times{\pp B}|/\mu_0$ as an estimate of the current helicity that only involves magnetic field measurements, as may be more readily available from spacecraft data.
We found that $S_H$ is 0.76, implying that $88\%$ of the energy transferred via $J_\parallel E_{\rm so, \parallel}$ occurs in the region with $H$ above its median value.

\begin{figure}[ht]
  \begin{center}
    \includegraphics[width=0.54\textwidth]{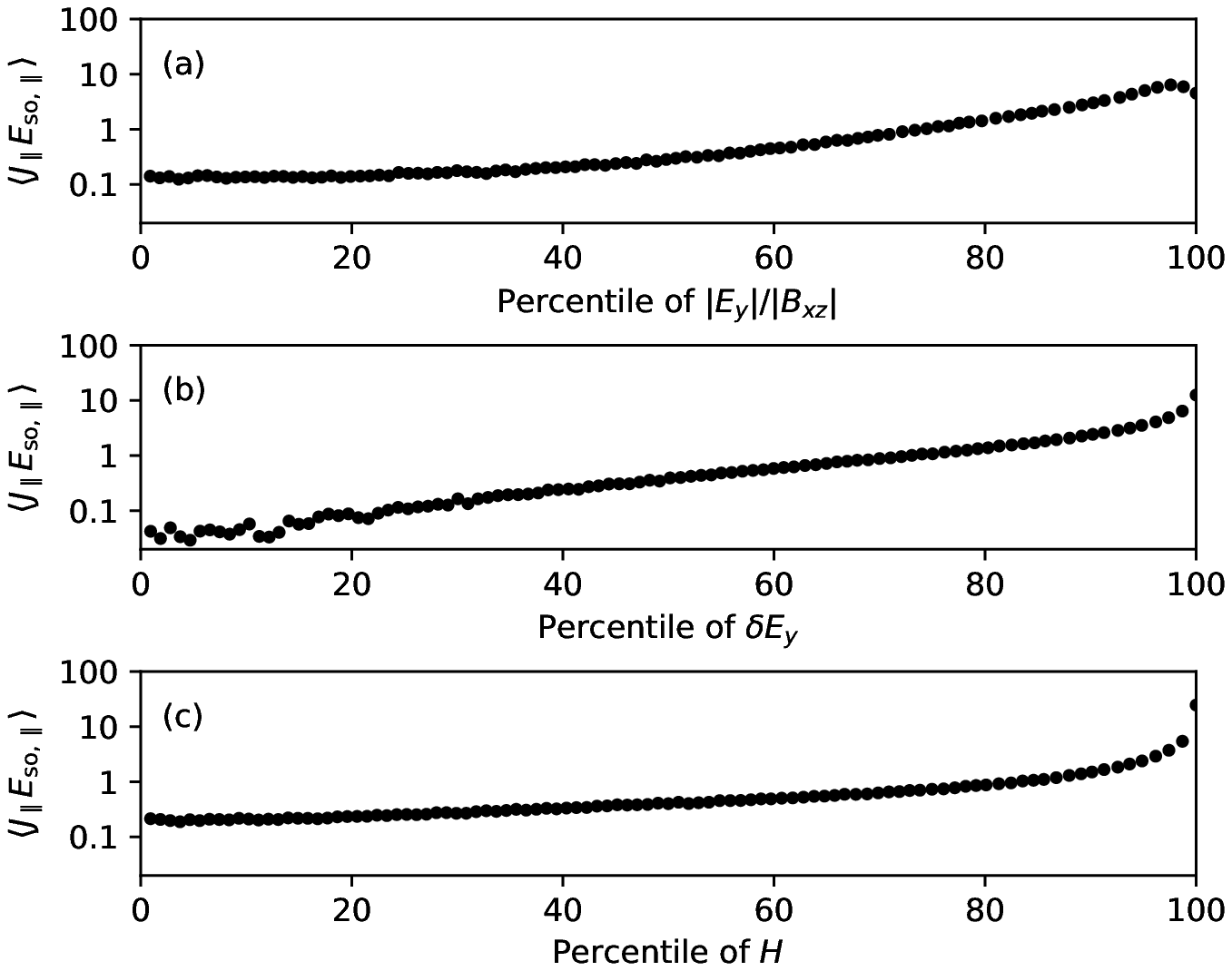}
    \caption{\small The average of $J_\parallel E_{\rm so \parallel}$ \add{normalized by its average value}  as a function of percentile of (a) $|E_y|/|B_{xz}|$, (b) $\delta E_y$, and (c) $H$ with $B_0/b_0=2$ for $t>\tau$.
    }
\label{fig:indc_his}
  \end{center}
\end{figure}

To demonstrate how $J_\parallel E_{\rm so, \parallel}$ is sensitive to the indicators, we numerically calculate the average value of $\langle J_\parallel E_{\rm so, \parallel}\rangle$ for each percentile of $X$. The percentile is calculated over the whole domain for $t>\tau$. 
Figure~\ref{fig:indc_his} shows $\langle J_\parallel E_{\rm so, \parallel}\rangle$ \add{normalized by its average value over the whole domain at $t>\tau$} as a function of the percentile of $\delta E_y$, $|E_y|/|B_{xz}|$ and $H$. The value of $\langle J_\parallel E_{\rm so, \parallel}\rangle$ becomes larger when the percentile of $|E_y|/|B_{xz}|$, $\delta E_y$ or $H$ is higher. The regions with one of $|E_y|/|B_{xz}|$ higher than $89$ percentile, $\delta E_y$ higher than $90$ percentile, or $H$ higher than $93$ percentile are responsible for $50\%$ of the energy conversion. 
Figures~\ref{fig:ej_locations}(b)-(d) show the regions selected by the values of  $|E_y|/|B_{xz}|$ (b), $\delta E_y$ (c) and $H$ (d) higher than their $90$ percentile by black coloring. The conversion sites selected by these indicators have either strongly positive or negative $J_\parallel E_{\rm so, \parallel}$.
The regions selected by these three indicators are quite different.
The regions selected by $|E_y|/|B_{xz}|$ are relatively contiguous. They are concentrated near the reconnection sites and away from the centers of magnetic islands. This is useful because, as noted above, the magnetic islands provide very little net energy conversion to the particles. $H$ selects both regions
near the reconnection sites and centers of magnetic islands, though the latter make very little contribution
to the net energy conversion.
The regions selected by $\delta E_y$ are scattered quite broadly but statistically concentrated around the reconnection sites.


\section{\label{discuss}Discussion and Summary}

We have run 2.5D simulations of decaying turbulence. There are two phases of the simulations that give two distinct mechanisms of magnetic energy conversion. These phases are significantly modified by the magnitude of the guide field. As $B_0$ is higher, plasma becomes less compressible as the magnetic field partially confines the particles. Another effect of higher $B_0$ is the larger amplitude of the irrotational electric field. This field component is not directly responsible for the magnetic energy conversion and we therefore focus on the solenoidal electric field when searching for indicators of energy conversion locations.

For $t<\tau$ or the island-forming phase, plasma current structures contract to form magnetic islands. In this phase, electron motions take energy out of the magnetic field but also lose a large amount of energy to ions via the irrotational electric field. While the in-plane magnetic field becomes weaker, the out-of-plane component becomes stronger at centers of magnetic islands by taking energy from the in-plane particle motions.
The current contraction obviously depends on the magnitude of the guide magnetic field.
As $B_0$ gets stronger, the fractional energy conversion in this phase is weaker, and the ratio of energy taken out of the in-plane magnetic field by electrons to ions becomes higher.
The irrotational electric field becomes stronger and is responsible for more energy transfer from electrons to ions. 
Both ions and electrons lose more energy to build strong $B_y$ at centers of magnetic islands. 

At $t>\tau$ or the decaying phase, the magnetic islands merge. During island merging, the magnetic reconnection inevitably occurs. As $B_0$ is higher, the particle parallel motion becomes dominant at taking energy out of the in-plane magnetic field~\citep{wan2012,makwana2017}, and while ions gain more energy from the magnetic field, they lose a large amount of energy to electrons via the irrotational electric field. Electrons eventually gains more energy in this phase.
Overall electrons become more effective at extracting the electromagnetic energy when $B_0$ is higher~\citep{shay2018}.  This result is similar to the beta effect on the particle heating~\citep{parashar2018}. 

The energy conversion is mainly mediated by the parallel component of the solenoidal electric field via $J_\parallel E_{\rm so,\parallel}$, which is highly localized. Three  particularly useful indicators of the regions of strong energy conversion are found. They are 1) the ratio of the out-of-plane electric field to the in-plane magnetic field \citep[related to the suggestion by][]{lapenta2021}, 2) the out-of-plane component of the non-ideal electric field, and 3) the magnitude of an estimate of current helicity. All of them select regions near the magnetic reconnection sites as locations of net energy conversion. Among them, $|E_y|/|B_{xz}|$ provides regions that are most contiguous and tied with multiple reconnection sites. We propose that any mechanisms involving these indicators should receive particular attention in order to gain more insight on how magnetic energy is converted to bulk and thermal motions in turbulence with a strong guide field. 

P.P. and D.R. would like to thank Thailand Science
Research and Innovation and Ministry of Higher Education, Science, Research and Innovation (Thailand) for support through grants RTA
6280002 and RGNS 63-045. This work was partly supported
by the International Atomic Energy Agency (IAEA) under
Contract No. 22785.
\nocite{*}

\bibliography{apssamp}

\begin{thebibliography}{}
\expandafter\ifx\csname natexlab\endcsname\relax\def\natexlab#1{#1}\fi
\providecommand{\url}[1]{\href{#1}{#1}}
\providecommand{\dodoi}[1]{doi:~\href{http://doi.org/#1}{\nolinkurl{#1}}}
\providecommand{\doeprint}[1]{\href{http://ascl.net/#1}{\nolinkurl{http://ascl.net/#1}}}
\providecommand{\doarXiv}[1]{\href{https://arxiv.org/abs/#1}{\nolinkurl{https://arxiv.org/abs/#1}}}

\bibitem[{{Bowers} {et~al.}(2008){Bowers}, {Albright}, {Yin}, {Bergen}, \&
  {Kwan}}]{bowers2008}
{Bowers}, K.~J., {Albright}, B.~J., {Yin}, L., {Bergen}, B., \& {Kwan},
  T.~J.~T. 2008, Physics of Plasmas, 15, 055703, \dodoi{10.1063/1.2840133}

\bibitem[{{Camporeale} {et~al.}(2018){Camporeale}, {Sorriso-Valvo}, {Califano},
  \& {Retin{\`o}}}]{camporeale2018}
{Camporeale}, E., {Sorriso-Valvo}, L., {Califano}, F., \& {Retin{\`o}}, A.
  2018, \prl, 120, 125101, \dodoi{10.1103/PhysRevLett.120.125101}

\bibitem[{{Chandran} {et~al.}(2010){Chandran}, {Li}, {Rogers}, {Quataert}, \&
  {Germaschewski}}]{chandran10}
{Chandran}, B. D.~G., {Li}, B., {Rogers}, B.~N., {Quataert}, E., \&
  {Germaschewski}, K. 2010, \apj, 720, 503, \dodoi{10.1088/0004-637X/720/1/503}

\bibitem[{{Chen} {et~al.}(2020){Chen}, {Shen}, {Gary}, {Reeves}, {Fleishman},
  {Yu}, {Guo}, {Krucker}, {Lin}, {Nita}, \& {Kong}}]{Chen2020}
{Chen}, B., {Shen}, C., {Gary}, D.~E., {et~al.} 2020, Nature Astronomy, 4,
  1140, \dodoi{10.1038/s41550-020-1147-7}

\bibitem[{{Comisso} \& {Sironi}(2018)}]{Comisso2018}
{Comisso}, L., \& {Sironi}, L. 2018, \prl, 121, 255101,
  \dodoi{10.1103/PhysRevLett.121.255101}

\bibitem[{{Dahlin} {et~al.}(2014){Dahlin}, {Drake}, \& {Swisdak}}]{Dahlin2014}
{Dahlin}, J.~T., {Drake}, J.~F., \& {Swisdak}, M. 2014, Physics of Plasmas, 21,
  092304, \dodoi{10.1063/1.4894484}

\bibitem[{{Daughton} {et~al.}(2011){Daughton}, {Roytershteyn}, {Karimabadi},
  {Yin}, {Albright}, {Bergen}, \& {Bowers}}]{daughton2011}
{Daughton}, W., {Roytershteyn}, V., {Karimabadi}, H., {et~al.} 2011, Nature
  Physics, 7, 539, \dodoi{10.1038/nphys1965}

\bibitem[{{De Pontieu} {et~al.}(2007){De Pontieu}, {McIntosh}, {Carlsson},
  {Hansteen}, {Tarbell}, {Schrijver}, {Title}, {Shine}, {Tsuneta}, {Katsukawa},
  {Ichimoto}, {Suematsu}, {Shimizu}, \& {Nagata}}]{pontieu07}
{De Pontieu}, B., {McIntosh}, S.~W., {Carlsson}, M., {et~al.} 2007, Science,
  318, 1574, \dodoi{10.1126/science.1151747}

\bibitem[{{Du} {et~al.}(2018){Du}, {Guo}, {Zank}, {Li}, \& {Stanier}}]{Du2018}
{Du}, S., {Guo}, F., {Zank}, G.~P., {Li}, X., \& {Stanier}, A. 2018, \apj, 867,
  16, \dodoi{10.3847/1538-4357/aae30e}

\bibitem[{{Du} {et~al.}(2020){Du}, {Zank}, {Li}, \& {Guo}}]{Du2020}
{Du}, S., {Zank}, G.~P., {Li}, X., \& {Guo}, F. 2020, \pre, 101, 033208,
  \dodoi{10.1103/PhysRevE.101.033208}

\bibitem[{{Fu} {et~al.}(2020){Fu}, {Guo}, {Li}, \& {Li}}]{Fu2020}
{Fu}, X., {Guo}, F., {Li}, H., \& {Li}, X. 2020, \apj, 890, 161,
  \dodoi{10.3847/1538-4357/ab6d68}

\bibitem[{{Gary} \& {Borovsky}(2004)}]{gary2004}
{Gary}, S.~P., \& {Borovsky}, J.~E. 2004, Journal of Geophysical Research
  (Space Physics), 109, A06105, \dodoi{10.1029/2004JA010399}

\bibitem[{{Gary} {et~al.}(2016){Gary}, {Hughes}, \& {Wang}}]{gary16}
{Gary}, S.~P., {Hughes}, R.~S., \& {Wang}, J. 2016, \apj, 816, 102,
  \dodoi{10.3847/0004-637X/816/2/102}

\bibitem[{{Guo} {et~al.}(2014){Guo}, {Li}, {Daughton}, \& {Liu}}]{Guo2014}
{Guo}, F., {Li}, H., {Daughton}, W., \& {Liu}, Y.-H. 2014, \prl, 113, 155005,
  \dodoi{10.1103/PhysRevLett.113.155005}

\bibitem[{{Guo} {et~al.}(2019){Guo}, {Li}, {Daughton}, {Kilian}, {Li}, {Liu},
  {Yan}, \& {Ma}}]{Guo2019}
{Guo}, F., {Li}, X., {Daughton}, W., {et~al.} 2019, \apjl, 879, L23,
  \dodoi{10.3847/2041-8213/ab2a15}

\bibitem[{{Guo} {et~al.}(2020{\natexlab{a}}){Guo}, {Li}, {Daughton}, {Li},
  {Kilian}, {Liu}, {Zhang}, \& {Zhang}}]{Guo2020b}
---. 2020{\natexlab{a}}, arXiv e-prints, arXiv:2008.02743.
\newblock \doarXiv{2008.02743}

\bibitem[{{Guo} {et~al.}(2015){Guo}, {Liu}, {Daughton}, \& {Li}}]{Guo2015}
{Guo}, F., {Liu}, Y.-H., {Daughton}, W., \& {Li}, H. 2015, \apj, 806, 167,
  \dodoi{10.1088/0004-637X/806/2/167}

\bibitem[{{Guo} {et~al.}(2020{\natexlab{b}}){Guo}, {Liu}, {Li}, {Li},
  {Daughton}, \& {Kilian}}]{Guo2020a}
{Guo}, F., {Liu}, Y.-H., {Li}, X., {et~al.} 2020{\natexlab{b}}, Physics of
  Plasmas, 27, 080501, \dodoi{10.1063/5.0012094}

\bibitem[{{Hughes} {et~al.}(2017){Hughes}, {Gary}, {Wang}, \&
  {Parashar}}]{hughes17}
{Hughes}, R.~S., {Gary}, S.~P., {Wang}, J., \& {Parashar}, T.~N. 2017, \apjl,
  847, L14, \dodoi{10.3847/2041-8213/aa8b13}

\bibitem[{Kida \& Orszag(1992)}]{kida1992}
Kida, S., \& Orszag, S.~A. 1992, Journal of Scientific Computing, 7, 1,
  \dodoi{10.1007/BF01060209}

\bibitem[{{Lapenta}(2021)}]{lapenta2021}
{Lapenta}, G. 2021, \apj, 911, 147, \dodoi{10.3847/1538-4357/abeb74}

\bibitem[{{Leamon} {et~al.}(1998){Leamon}, {Matthaeus}, {Smith}, \&
  {Wong}}]{leamon98}
{Leamon}, R.~J., {Matthaeus}, W.~H., {Smith}, C.~W., \& {Wong}, H.~K. 1998,
  \apjl, 507, L181, \dodoi{10.1086/311698}

\bibitem[{{Li} {et~al.}(2019{\natexlab{a}}){Li}, {Guo}, \& {Li}}]{Li2019}
{Li}, X., {Guo}, F., \& {Li}, H. 2019{\natexlab{a}}, \apj, 879, 5,
  \dodoi{10.3847/1538-4357/ab223b}

\bibitem[{{Li} {et~al.}(2018){Li}, {Guo}, {Li}, \& {Birn}}]{li2018}
{Li}, X., {Guo}, F., {Li}, H., \& {Birn}, J. 2018, \apj, 855, 80,
  \dodoi{10.3847/1538-4357/aaacd5}

\bibitem[{{Li} {et~al.}(2017){Li}, {Guo}, {Li}, \& {Li}}]{li2017}
{Li}, X., {Guo}, F., {Li}, H., \& {Li}, G. 2017, \apj, 843, 21,
  \dodoi{10.3847/1538-4357/aa745e}

\bibitem[{{Li} {et~al.}(2019{\natexlab{b}}){Li}, {Guo}, {Li}, {Stanier}, \&
  {Kilian}}]{Li2019b}
{Li}, X., {Guo}, F., {Li}, H., {Stanier}, A., \& {Kilian}, P.
  2019{\natexlab{b}}, \apj, 884, 118, \dodoi{10.3847/1538-4357/ab4268}

\bibitem[{{Makwana} {et~al.}(2017){Makwana}, {Li}, {Guo}, \&
  {Li}}]{makwana2017}
{Makwana}, K., {Li}, H., {Guo}, F., \& {Li}, X. 2017, in Journal of Physics
  Conference Series, Vol. 837, Journal of Physics Conference Series, 012004

\bibitem[{{Montgomery}(1982)}]{montgomery82}
{Montgomery}, D. 1982, Physica Scripta Volume T, 2A, 83,
  \dodoi{10.1088/0031-8949/1982/T2A/009}

\bibitem[{{Parashar} {et~al.}(2018){Parashar}, {Matthaeus}, \&
  {Shay}}]{parashar2018}
{Parashar}, T.~N., {Matthaeus}, W.~H., \& {Shay}, M.~A. 2018, \apjl, 864, L21,
  \dodoi{10.3847/2041-8213/aadb8b}

\bibitem[{{Parashar} {et~al.}(2015){Parashar}, {Salem}, {Wicks}, {Karimabadi},
  {Gary}, \& {Matthaeus}}]{parashar2015}
{Parashar}, T.~N., {Salem}, C., {Wicks}, R.~T., {et~al.} 2015, Journal of
  Plasma Physics, 81, 905810513, \dodoi{10.1017/S0022377815000860}

\bibitem[{{Parker}(1988)}]{parker88}
{Parker}, E.~N. 1988, \apj, 330, 474, \dodoi{10.1086/166485}

\bibitem[{{Perri} {et~al.}(2012){Perri}, {Goldstein}, {Dorelli}, \&
  {Sahraoui}}]{perri2012}
{Perri}, S., {Goldstein}, M.~L., {Dorelli}, J.~C., \& {Sahraoui}, F. 2012,
  \prl, 109, 191101, \dodoi{10.1103/PhysRevLett.109.191101}

\bibitem[{{Pongkitiwanichakul} {et~al.}(2015){Pongkitiwanichakul}, {Cattaneo},
  {Boldyrev}, {Mason}, \& {Perez}}]{peera2015}
{Pongkitiwanichakul}, P., {Cattaneo}, F., {Boldyrev}, S., {Mason}, J., \&
  {Perez}, J.~C. 2015, \mnras, 454, 1503, \dodoi{10.1093/mnras/stv2008}

\bibitem[{{Sahraoui} {et~al.}(2009){Sahraoui}, {Goldstein}, {Robert}, \&
  {Khotyaintsev}}]{sahraoui2009}
{Sahraoui}, F., {Goldstein}, M.~L., {Robert}, P., \& {Khotyaintsev}, Y.~V.
  2009, \prl, 102, 231102, \dodoi{10.1103/PhysRevLett.102.231102}

\bibitem[{{Shay} {et~al.}(2018){Shay}, {Haggerty}, {Matthaeus}, {Parashar},
  {Wan}, \& {Wu}}]{shay2018}
{Shay}, M.~A., {Haggerty}, C.~C., {Matthaeus}, W.~H., {et~al.} 2018, Physics of
  Plasmas, 25, 012304, \dodoi{10.1063/1.4993423}

\bibitem[{{TenBarge} \& {Howes}(2013)}]{tenBarge2013}
{TenBarge}, J.~M., \& {Howes}, G.~G. 2013, \apjl, 771, L27,
  \dodoi{10.1088/2041-8205/771/2/L27}

\bibitem[{{van Ballegooijen}(1986)}]{ballegooijen86}
{van Ballegooijen}, A.~A. 1986, \apj, 311, 1001, \dodoi{10.1086/164837}

\bibitem[{{Wan} {et~al.}(2015){Wan}, {Matthaeus}, {Roytershteyn}, {Karimabadi},
  {Parashar}, {Wu}, \& {Shay}}]{wan2015}
{Wan}, M., {Matthaeus}, W.~H., {Roytershteyn}, V., {et~al.} 2015, \prl, 114,
  175002, \dodoi{10.1103/PhysRevLett.114.175002}

\bibitem[{{Wan} {et~al.}(2016){Wan}, {Matthaeus}, {Roytershteyn}, {Parashar},
  {Wu}, \& {Karimabadi}}]{wan2016}
---. 2016, Physics of Plasmas, 23, 042307, \dodoi{10.1063/1.4945631}

\bibitem[{{Wan} {et~al.}(2012){Wan}, {Matthaeus}, {Karimabadi}, {Roytershteyn},
  {Shay}, {Wu}, {Daughton}, {Loring}, \& {Chapman}}]{wan2012}
{Wan}, M., {Matthaeus}, W.~H., {Karimabadi}, H., {et~al.} 2012, \prl, 109,
  195001, \dodoi{10.1103/PhysRevLett.109.195001}

\bibitem[{{Wu} {et~al.}(2013){Wu}, {Wan}, {Matthaeus}, {Shay}, \&
  {Swisdak}}]{wu13}
{Wu}, P., {Wan}, M., {Matthaeus}, W.~H., {Shay}, M.~A., \& {Swisdak}, M. 2013,
  \prl, 111, 121105, \dodoi{10.1103/PhysRevLett.111.121105}

\bibitem[{{Yang} {et~al.}(2021){Yang}, {Wan}, {Matthaeus}, \&
  {Chen}}]{yang2021}
{Yang}, Y., {Wan}, M., {Matthaeus}, W.~H., \& {Chen}, S. 2021, Journal of Fluid
  Mechanics, 916, A4, \dodoi{10.1017/jfm.2021.199}

\bibitem[{{Yang} {et~al.}(2019){Yang}, {Wan}, {Matthaeus}, {Sorriso-Valvo},
  {Parashar}, {Lu}, {Shi}, \& {Chen}}]{yang2019}
{Yang}, Y., {Wan}, M., {Matthaeus}, W.~H., {et~al.} 2019, \mnras, 482, 4933,
  \dodoi{10.1093/mnras/sty2977}

\bibitem[{{Yang} {et~al.}(2017{\natexlab{a}}){Yang}, {Matthaeus}, {Parashar},
  {Haggerty}, {Roytershteyn}, {Daughton}, {Wan}, {Shi}, \& {Chen}}]{Yang2017}
{Yang}, Y., {Matthaeus}, W.~H., {Parashar}, T.~N., {et~al.} 2017{\natexlab{a}},
  Physics of Plasmas, 24, 072306, \dodoi{10.1063/1.4990421}

\bibitem[{{Yang} {et~al.}(2017{\natexlab{b}}){Yang}, {Matthaeus}, {Parashar},
  {Wu}, {Wan}, {Shi}, {Chen}, {Roytershteyn}, \& {Daughton}}]{Yang2017PRE}
---. 2017{\natexlab{b}}, \pre, 95, 061201, \dodoi{10.1103/PhysRevE.95.061201}

\bibitem[{{Zhang} {et~al.}(2015){Zhang}, {Chen}, {B{\"o}ttcher}, {Guo}, \&
  {Li}}]{Zhang2015}
{Zhang}, H., {Chen}, X., {B{\"o}ttcher}, M., {Guo}, F., \& {Li}, H. 2015, \apj,
  804, 58, \dodoi{10.1088/0004-637X/804/1/58}

\bibitem[{{Zhang} {et~al.}(2020){Zhang}, {Li}, {Giannios}, {Guo}, {Liu}, \&
  {Dong}}]{Zhang2020}
{Zhang}, H., {Li}, X., {Giannios}, D., {et~al.} 2020, \apj, 901, 149,
  \dodoi{10.3847/1538-4357/abb1b0}

\bibitem[{{Zhang} {et~al.}(2018){Zhang}, {Li}, {Guo}, \&
  {Giannios}}]{Zhang2018}
{Zhang}, H., {Li}, X., {Guo}, F., \& {Giannios}, D. 2018, \apjl, 862, L25,
  \dodoi{10.3847/2041-8213/aad54f}

\bibitem[{{Zhdankin} {et~al.}(2019){Zhdankin}, {Uzdensky}, {Werner}, \&
  {Begelman}}]{zhdankin2019}
{Zhdankin}, V., {Uzdensky}, D.~A., {Werner}, G.~R., \& {Begelman}, M.~C. 2019,
  \prl, 122, 055101, \dodoi{10.1103/PhysRevLett.122.055101}

\end{thebibliography}

\end{document}